\documentclass[onecolumn, tighten]{aastex631}
\usepackage{pdfpages}
\usepackage{nicefrac}
\usepackage{microtype}
\usepackage{booktabs}
\usepackage{rotating}
\usepackage{xcolor}
\usepackage[utf8]{inputenc}
\usepackage{amsmath}
\usepackage{etoolbox} 

\makeatletter
\patchcmd{\@outputpage@head}{\@ifx{\LS@rot\@undefined}{}{\LS@rot}}{}{}{}
\makeatother

\DeclareUnicodeCharacter{2212}{\ensuremath{-}}

\received{?}
\revised{?}
\accepted{?}
\submitjournal{Frontiers}
\shorttitle{Ionospheric Joule Heating}
\shortauthors{Bagheri and Lopez}

\begin{document}

\title{Comparison of Empirical Models of Ionospheric Heating to Global Simulations}

\author{Fatemeh Bagheri}
\affil{Department of Physics, University of Texas at Arlington, Arlington, TX, USA}
\correspondingauthor{Fatemeh Bagheri}
\email{fatemeh.bagheri@uta.edu}

\author{Ramon E. Lopez}
\affil{Department of Physics, University of Texas at Arlington, Arlington, TX, USA}

\begin{abstract}


\noindent Intense currents produced during geomagnetic storms dissipate energy in the ionosphere through Joule heating. This dissipation has significant space weather effects, and thus it is important to determine the ability of physics-based simulations to replicate real events quantitatively. Several empirical models estimate Joule heating based on ionospheric currents using the AE index. In this study, we select 11 magnetic storm simulations from the CCMC database and compare the integrated Joule heating in the simulations with the results of empirical models. We also use the SWMF global magnetohydrodynamic simulations for 12 storms to reproduce the correlation between the simulated AE index and simulated Joule heating. We find that the scale factors in the empirical models are half what is predicted by the SWMF simulations.

\end{abstract}

\section{Introduction}

Dayside merging and nightside reconnection produce plasma flow in the ionosphere, which can be intense and steady during a geomagnetic storm. The flow means that there is an electric field in Earth's reference frame, and this electric field drives auroral electrojet currents that close the Birkeland currents driven by reconnection. This current dissipates energy in the ionosphere through frictional heating, which is generally referred to as Joule heating, although actual electromagnetic Joule heating should be calculated in the plasma frame \citep{vasyliunas2005meaning}. The energy dissipated through the Joule heating process is the second most important energy sink after the ring current \citep{akasofu1981energy} or even sometimes the most important \citep{harel1981quantitative, lu1998global}. Thermospheric responses to Joule heating during magnetic storms can be quite significant \citep{deng2018possible}. The ionospheric Joule dissipation heats the ionosphere and thermosphere so they can expand upward. These upward expansions can produce increased ionospheric ion outflow and satellite drag. The effect of satellite drag and the changes in the drag produced by space weather is an important effect that needs to be quantified \citep{doornbos2006modelling}. Thus it is essential to understand how well the Joule heating produced by physics-based global simulations of magnetic storms compares to empirical estimates based on observations if such physics-based simulations are to be used for space weather prediction. 
\\
\\
Studies often use empirical models of electric fields and conductivities to estimate Joule heating. These models typically do not represent the variability of electric fields, currents, and conductivities about the average. The contribution of electric-field variance to total Joule heating and its thermosphere/ionosphere effects can be substantial \citep{richmond2021joule}. Therefore, to understand the energy transfer during geomagnetic storms and the coupling mechanism between the solar wind and the thermosphere-ionosphere-magnetosphere system, it is necessary to estimate the Joule heating rate accurately \citep{richmond2000upper}. 
\\
\\
Dissipation of energy through Joule heating is due to the current parallel to the electric field ($U_J = \vec{J}. \vec{E}$). Hence the height-integrated Joule heating can be expressed as a function of Pedersen conductivity in the reference frame of the neutrals,
\begin{equation}
U_J = \int \sigma_P(h) (E + U \times B)^2 dh
\end{equation}
where $\sigma_P$ is the Pedersen conductivity, $U$ is the neutral winds, and $B$ is the magnetic field. Calculating the Joule heating rate requires Pedersen conductivity and electric field measurements over the entire polar region. However, there is still a challenge to monitor these quantities continuously over the entire high-latitude region. Ionospheric electric fields and conductivities could be directly measured \emph{locally} by using rocket-borne instrumentation or more widely with incoherent scatter radar. Therefore several empirical models have been developed using geomagnetic indices such as Kp, AE, or AL to estimate the first approximation measure of the global Joule heating rates \citep{ahn1983joule, foster1983joule, baumjohann1984hemispherical}. For example, \citet{baumjohann1984hemispherical} assumed that the height‐integrated ionospheric conductivity simulates substorm activity in the AE index \citep{spiro1982precipitating, zhou2011joule}. However, these empirical models usually use small data sets and are based on assumptions that may not always be valid. For instance, \citet{baumjohann1984hemispherical} used the inversion method discussed by \citep{kamide1981estimation}, however, that technique may not always yield the best results and reflect the actual ionospheric electric fields and currents, a fact that was noted by the authors themselves. The empirical model of \citet{weimer2005improved} also solves for the ionospheric potential but uses a much larger database for its solution. However, for the purposes of this paper, we will use simple AE-based empirical formulations.  This will allow us to determine if the global simulated Joule heating is related to the simulated AE in a manner similar to the empirical relationship between the Joule heating calculated from observation and the observed AE.
\\ 
\\
To estimate the energy dissipated through the Joule heating process, one can use global Magneto-Hydro-Dynamic (MHD) simulations. Such models have been used for many years to simulate storms and substorms and investigate the transfer of energy in the geospace system \citep[e. g.,][]{lopez1998simulation,lopez2011role}. \citet{palmroth2004ionospheric} used the global MHD simulation code GUMICS-4 and found that the temporal variation of the Joule heating during substorms is well correlated with a commonly used AE-based empirical model, whereas, during the storm period, the simulated Joule heating was different from the empirical model. Following that study, in this paper, we use the Space Weather Modeling Framework (SWMF) MHD simulation for 12 storm events that had already been simulated with the results available at the Community Coordinated Modeling Center (CCMC) to compare the Joule heating resulting from simulations with empirical models. The
SWMF simulation combines numerical models of the Inner Heliosphere, Solar Energetic Particles, Global Magnetosphere, Inner Magnetosphere, Radiation Belt, Ionosphere, and Upper Atmosphere into a parallel, high-performance model \citep{toth2005space}. Two versions of the SWMF model are used on CCMC. All simulations selected in this paper are the version of v20180525.

\section{Correlations between SWMF Simulations and Observations, using the SME Index}

In this section, we compare three empirical models of Joule heating with the output of SMWF simulation. We select 11 magnetic storms with $Dst^* \leq -50$ nT for the period between 2010 and 2020 from the storm sample provided in \citep{bagheri2022solar}. The information on the SWMF simulations of these storms at the CCMC is listed in Table \ref{t:18JHvsSME} (for more information, see Appendix A). We calculate the Joule heating using three empirical formulas that relate the Joule heating to the AE index:
\begin{itemize}
\item Model 1: $U_{JH} (GW) = 0.32 AE$  \citep{baumjohann1984hemispherical},
\item Model 2: $U_{JH} (GW) = 0.28 AE + 0.9$ \citep{ostgaard2002energy, ostgaard2002relation},
\item Model 3: $U_{JH} (GW) = 0.23 AE$ \citep{kalafatoglu2018investigating}.
\end{itemize}

\noindent The AE index \citep{nose2015geomagnetic} is produced at a 1-min cadence using data from up to 12 magnetometer stations at latitudes that correspond to the average location of the auroral oval. SuperMAG now produces SME, an equivalent to AE, at a 1-min cadence \citep{gjerloev2012supermag}. SME is the difference between upper (SMU) and lower (SML) indices. SMU and SML are based on the H-component measured at stations in the latitudes of the auroral oval, with baseline removal carried out. The difference between AE and SME is the number of stations used in their derivation. While AE uses (at maximum) 12 stations, the number of stations used to derive SME is roughly an order of magnitude larger and the stations cover a broader range of latitude. Using the 1-min SME data from SuperMag in equation (2) instead of AE, we can better estimate the energy dissipated through Joule heating for each storm since the SME index has better coverage than AE, particularly at lower latitudes where intense electrojets can be found during magnetic storms because of the expanded polar cap.
Furthermore, we use data from Active Magnetosphere and Planetary Electrodynamics Response Experiment (AMPERE) to measure the strength of Birkeland (Field-Aligned Currents (FACs)) current during each storm. AMPERE produces global maps of the Birkeland current using magnetometer data from over 70 satellites in the Iridium network with a cadence of 2 minutes \citep{anderson2000sensing}.

\begin{table}[h]
\centering
\begin{tabular}{lccc}
\hline
Event Date &Time of Main Phase& Run Number& Grid\\
\hline
2 May 2010	&10-19 UT&		Pelin\_Erdemir\_021419\_1	& 1 M\\
28 May 2011	&06-12 UT& 		Haonan\_Wu\_071818\_1		& 18.5 M \\
5-6 August 2011	&19-04 UT	&Sean\_Blake\_042619\_4	& 1 M\\
26 Septemebr 2011	&15-22 UT	&	Pauline\_Dredger\_082321\_1	& 1 M \\
22 January 2012	&07-22 UT&	Diptiranjan\_Rout\_060919\_1	& 1 M\\
24 January 2012	&15-20 UT&	Joaquin\_Diaz\_011221\_1	& 1 M\\
17 June 2012	&00-14 UT&	Yihua\_zheng\_113015\_1	& 1 M\\
15 July 2012	&00-19 UT&		Antti\_Lakka\_070918\_1	& 1 M\\
8-9 October 2012 &19-08 UT&	Sean\_Blake\_042619\_7	& 1 M\\
13-14 November 2012	&23-08 UT& 	Siyuan\_Wu\_120519\_2	& 1 M\\
9 September 2015	&00-13 UT& Lei\_Cai\_071720\_1	& 2 M\\
\hline
\end{tabular}
\caption{Information on SWMF simulations on CCMC used to compare the simulated Joule heating and three empirical models. All selected simulations are real storm events and the version of v20180525.}\label{t:18JHvsSME}
\end{table}

\begin{figure}[ht!]
\includegraphics[width=0.5\columnwidth]{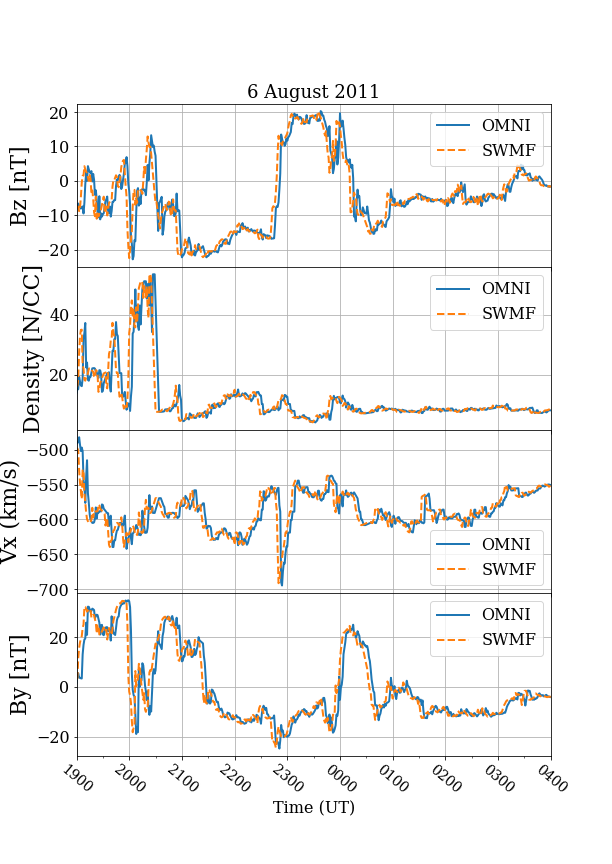}
\includegraphics[width=0.5\columnwidth]{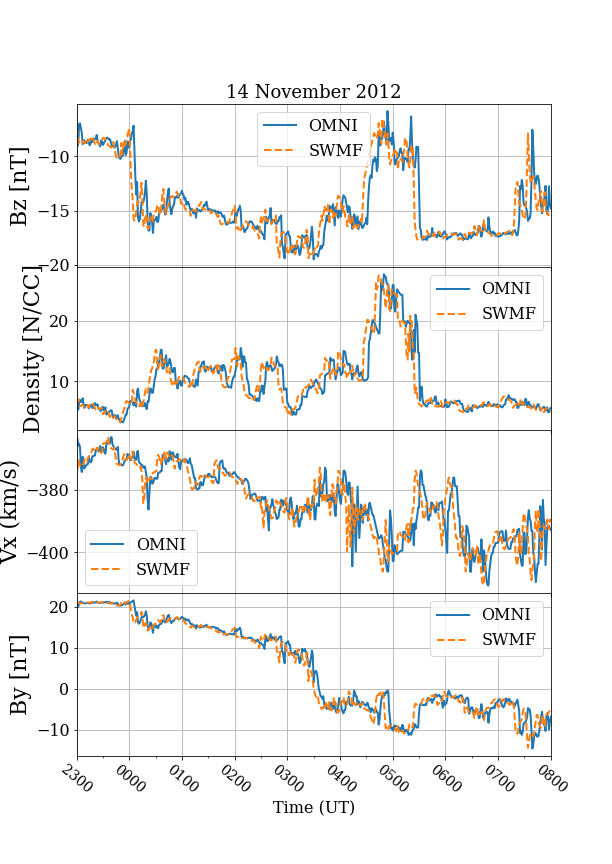}
\caption{\emph{left:}The solar wind data from OMNI and SWMF input for the magnetic storm on 6 August 2011. The simulation input and OMNI data are highly correlated with a correlation coefficient of 0.99 with a lagtime of 4 minutes. \emph{right:} The solar wind data from OMNI and SWMF input for the magnetic storm on 14 November 2012. The simulation input and OMNI data are highly correlated with a correlation coefficient of 0.99 with a lagtime of 6 minutes.}
\label{fig:GoodExample}
\end{figure}

\noindent We compare the solar wind input for the simulations to the 1-minute OMNI data provided by CDAWeb for each event. We only use the simulations whose inputs are in perfect agreement with OMNI data. Figure \ref{fig:GoodExample} illustrates two examples of storms where the solar wind data from OMNI is the same as the simulation input. This is not the case for some storms in the CCMC database, and there can be a substantial difference between the input for the simulations and the actual OMNI data during the event.\\

\begin{figure}[ht!]
\centering
\includegraphics[width=0.5\columnwidth]{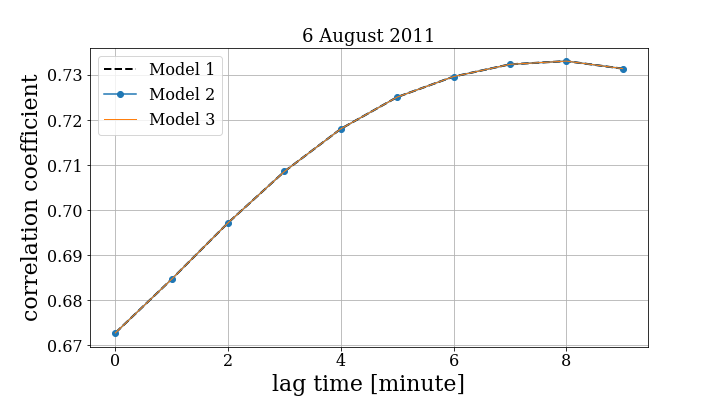}
\includegraphics[width=0.5\columnwidth]{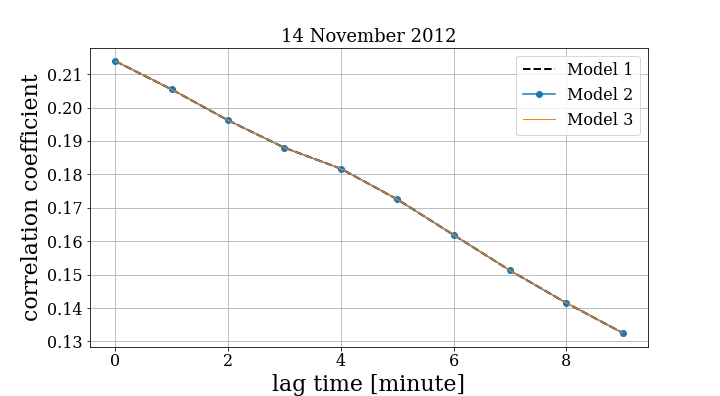}
\caption{The correlation coefficient of Joule heating resulted from three empirical models and SWMF simulations as a function of lagtime during the main phase of \emph{left:} the storm happened on 6 August 2011, and,
\emph{right:} the storm happened on 14 November 2012.}
\label{fig:JHexample}
\end{figure}

\noindent We find that good agreement of the simulation input with the OMNI data does not necessarily result in a good correlation between empirical and SWMF Joule heating. For instance, although in both cases shown in Figure \ref{fig:GoodExample}, the inputs of SWMF simulations of the storms are consistent with the OMNI data, in the first event (6 August 2011), the resulting Joule heating is highly correlated with all the three empirical models, whereas in the second event (14 November 2012) they are not (Figure \ref{fig:JHexample}). The lagtime between OMNI data and SWMF inputs in Figure \ref{fig:GoodExample} and \ref{fig:JHexample} is because OMNI data report the value of the solar wind parameters as they projected at the Bow shock region, while SMWF input data are projected at 32 RE.

\noindent We calculate the Pearson correlation coefficient between each empirical model of Joule heating and SWMF-simulated Joule heating for all storms. Moreover, since each storm has a different duration (i.e., different sample size), we calculated the standard error for Pearson correlation using Fisher's r-to-z transformation method, which results in asymmetrical confidence intervals. Furthermore, for each storm, we find the best linear fit of the simulated Joule heating as a function of Joule heating from empirical models. We summarize our results in Table \ref{t:resultsJHs}. All three empirical models have the same Pearson correlation coefficient up to 5 decimal digits. However, Model 2: $U_{JH} (GW) = 0.28 AE + 0.9$ [\citep{ostgaard2002energy, ostgaard2002relation}] has the greatest slopes in all events, and thus it fits better to the simulations.\\

\begin{table}[ht!]
\centering
\begin{tabular}{lcccc}
\hline
Event Date &c.c.& slope for Model 1 &slope for Model 2 & slope for Model 3\\
\hline
2 May 2010	&$0.66_{- 0.10}^{+0.08}$&		$0.45 \pm 0.04$     &$0.63 \pm 0.06$    &$0.54 \pm 0.05$\\
28 May 2011	&$0.47_{- 0.09}^{+0.08}$& 	    $0.13 \pm 0.01$     &$0.18 \pm 0.02$    &$0.15 \pm 0.01$\\
5-6 August 2011	&$0.73_{- 0.04}^{+0.03}$&	$0.70 \pm 0.02$     &$0.97 \pm 0.03$    &$0.83 \pm 0.03$\\
26 September 2011	&$0.35_{- 0.11}^{+0.10}$&	$0.35 \pm 0.06$     &$0.49 \pm 0.08$    &$0.42 \pm 0.07$\\
22 January 2012	&$0.61_{- 0.09}^{+0.07}$&	$0.77 \pm 0.06$     &$1.07 \pm 0.09$    &$0.91 \pm 0.08$\\
24 January 2012	&$0.46_{- 0.09}^{+0.08}$&	$0.50 \pm 0.05$     &$0.69 \pm 0.07$    &$0.59 \pm 0.06$\\
17 June 2012	&$0.49_{- 0.11}^{+0.09}$&	$0.48 \pm 0.05$     &$0.56 \pm 0.07$    &$0.48 \pm 0.06$\\
15 July 2012	&$0.76_{- 0.05}^{+0.04}$&	$0.67 \pm 0.03$     &$0.93 \pm 0.04$    &$0.79 \pm 0.04$\\
8-9 October 2012 &$0.55_{- 0.05}^{+0.05}$&	$0.22 \pm 0.01$     &$0.31 \pm 0.01$    &$0.26 \pm 0.01$\\
13-14 November 2012	&$0.21_{- 0.08}^{+0.07}$& 	$0.22 \pm 0.04$     &$0.30 \pm 0.06$    &$0.26 \pm 0.05$\\
9 September 2015	&$0.80_{- 0.19}^{+0.10}$&   $0.21 \pm 0.03$     &$0.30 \pm 0.04$    &$0.25 \pm 0.03$\\
\hline
\end{tabular}
\caption{Correlation coefficients between the simulated Joule heating and Joule heating from empirical models. All three empirical models have the same correlation coefficient of up to 5 decimals. Moreover, the slopes of the best linear fits of the simulated Joule heating as a function of Joule heating from empirical models are reported in columns 3-5.}\label{t:resultsJHs}
\end{table}
\noindent Additionally, we investigate the Pearson correlation between the AMPERE Birkeland current and the Birkeland current in the simulations. Similar to the previous Joule heating calculations, we calculated the standard error for Pearson correlation using Fisher's r-to-z transformation method. We also find the best linear fits of simulated and the AMPERE Birkeland currents. As represented in Table \ref{t:resultsCurrents}, SWMF simulations predict a smaller amount of Birkeland currents for all events in this study, approximately by a factor of $1/3$ for the events with $c.c. \geq 0.8$.  

\begin{figure}[ht!]
\centering
\includegraphics[width=10cm]{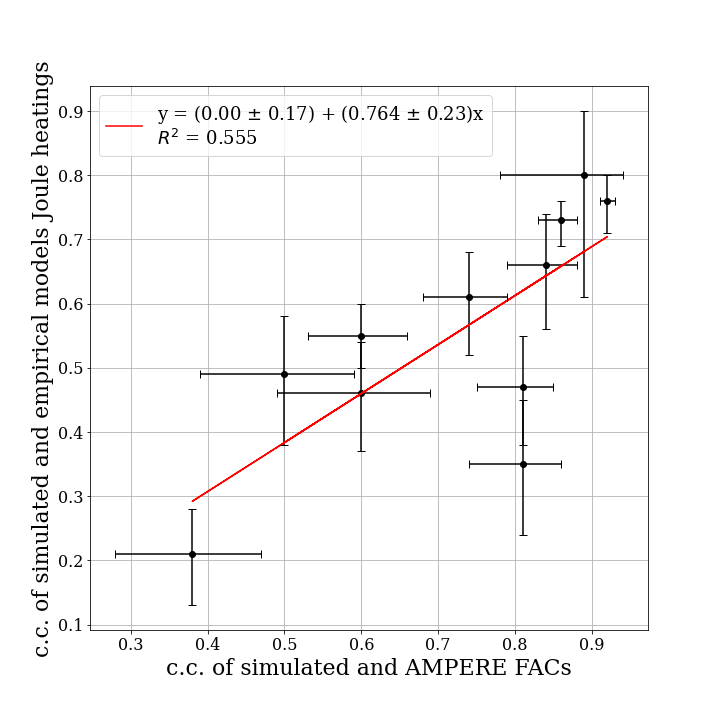}
\caption{Pearson correlation coefficient between the empirical and simulated Joule heating as a function of the correlation coefficient between the empirical and simulated Birkeland currents for the storms in our sample. All the standard errors for Pearson correlation were calculated by using Fisher's r-to-z transformation method.}
\label{fig:CC_JH_vs_I}
\end{figure}

\begin{table}[ht!]
\centering
\begin{tabular}{lccc}
\hline
Event Date &c.c& best fit slope \\
\hline
2 May 2010	&$0.84_{- 0.05}^{+0.04}$&		$0.31 \pm 0.01$\\
28 May 2011	&$0.81_{- 0.06}^{+0.04}$& 	    $0.42 \pm 0.02$\\
5-6 August 2011	&$0.86_{- 0.03}^{+0.02}$&	$0.41 \pm 0.01$\\
26 September 2011	&$0.81_{- 0.07}^{+0.05}$&	$0.31 \pm 0.02$\\
22 January 2012	&$0.74_{- 0.06}^{+0.05}$&	$0.19 \pm 0.01$\\
24 January 2012	&$0.60_{- 0.11}^{+0.09}$&	$0.31 \pm 0.03$\\
17 June 2012	&$0.50_{- 0.11}^{+0.09}$&	$0.17 \pm 0.02$\\
15 July 2012	&$0.92_{- 0.01}^{+0.01}$&	$0.49 \pm 0.01$\\
8-9 October 2012 &$0.60_{- 0.07}^{+0.06}$&	$0.22 \pm 0.01$\\
13-14 November 2012	&$0.38_{- 0.10}^{+0.09}$& 	$0.24 \pm 0.03$\\
9 September 2015	&$0.89_{- 0.11}^{+0.05}$&   $0.14 \pm 0.01$\\
\hline
\end{tabular}
\caption{Correlation coefficients and the slopes of best linear fits of the simulated FAC and AMPERE data.}\label{t:resultsCurrents}
\end{table}
\noindent As represented in Figure \ref{fig:CC_JH_vs_I}  we find that the correlation coefficient between Joule heatings (simulated and empirical) increases as the correlation coefficient of Birkeland currents (simulated and observed by AMPERE) increases. This result corroborates results in \citep{robinson2021determination}. They showed the SME index could be accurately deduced from AMPERE data with a correlation coefficient of $0.84$. In other words, if SWMF simulation predicts the Birkeland currents correctly, then the Joule heating would be simulated consistently with observations. This is not surprising for two reasons. First, in the simulation results, agreeing with the observed Birkeland current intensity, one can have confidence that the simulation accurately represented the solar wind-magnetosphere interaction and that other simulation features would also bear a reasonable resemblance to reality. Moreover, since the auroral electrojets are the ionospheric closure currents for the Birkeland currents (as represented by the correlation between the simulated SME and simulated Birkeland current), getting the Birkeland currents right will mean that the SME will also be (more or less) correct, at least in terms of the variations if not the absolute magnitude. This result can be used to identify periods when the real-time SMWF simulation of Joule heating is accurate during the geomagnetic storms by calculating a running correlation between the SWMF results and Birkeland current observations. 

\section{Correlations between AE and Joule Heating using SWMF Simulations}

\begin{figure}[ht!]
\centering
{\includegraphics[clip,width=0.3\columnwidth]{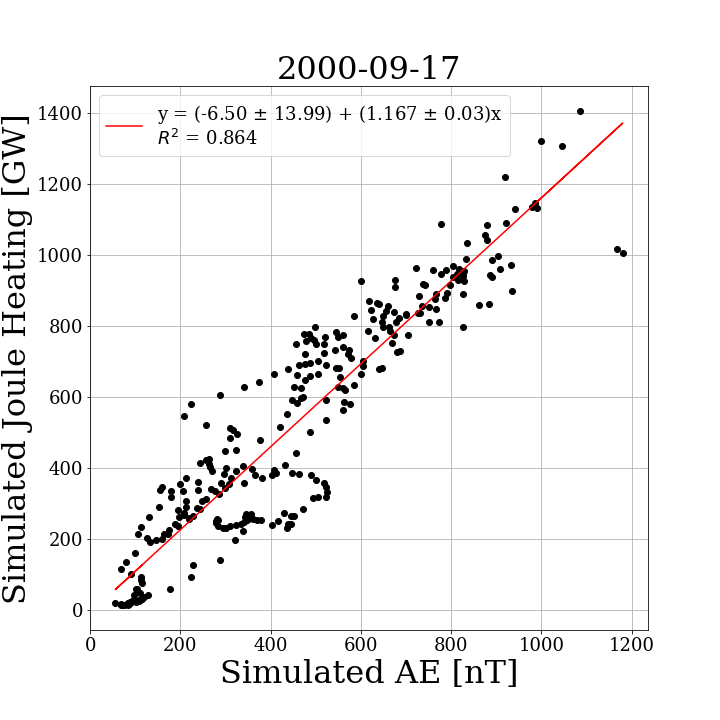}}%
{\includegraphics[clip,width=0.3\columnwidth]{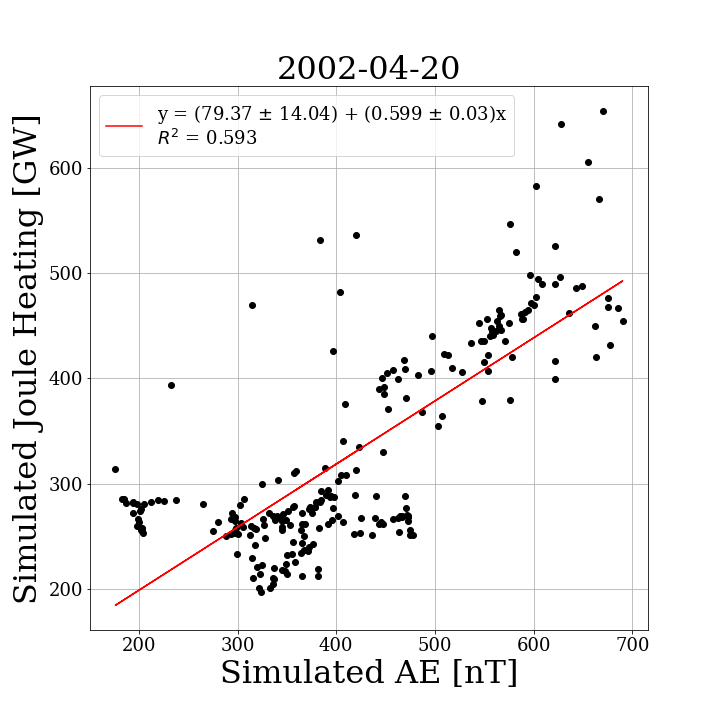} }%
{\includegraphics[clip,width=0.3\columnwidth]{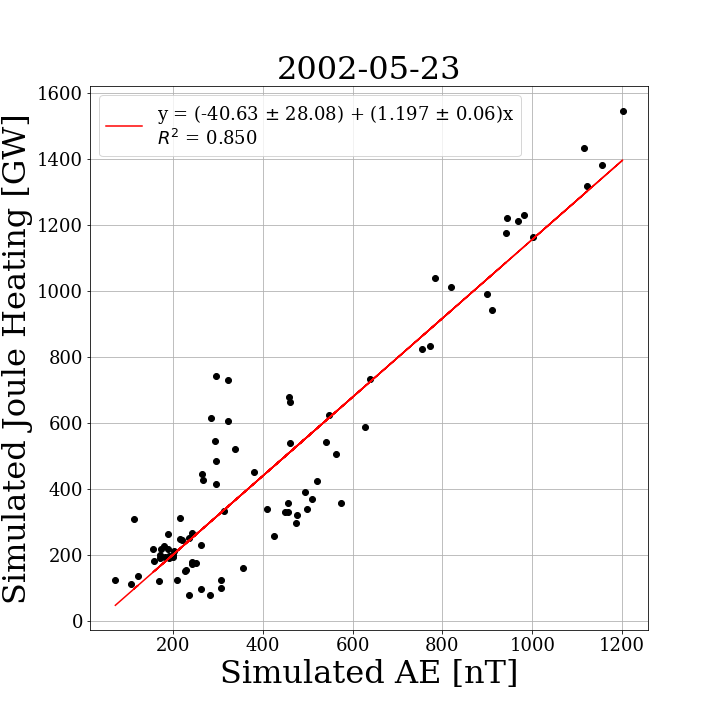}}\\%
{\includegraphics[clip,width=0.3\columnwidth]{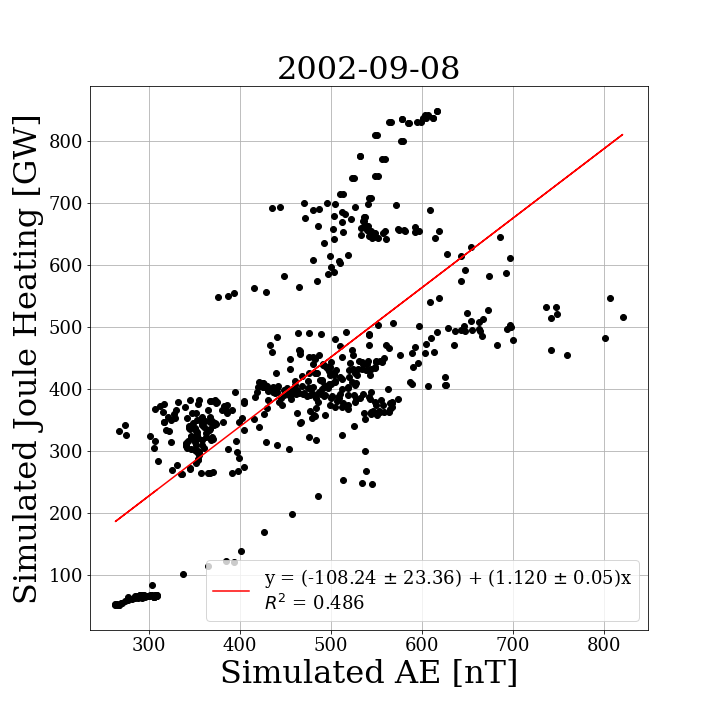} }%
{\includegraphics[clip,width=0.3\columnwidth]{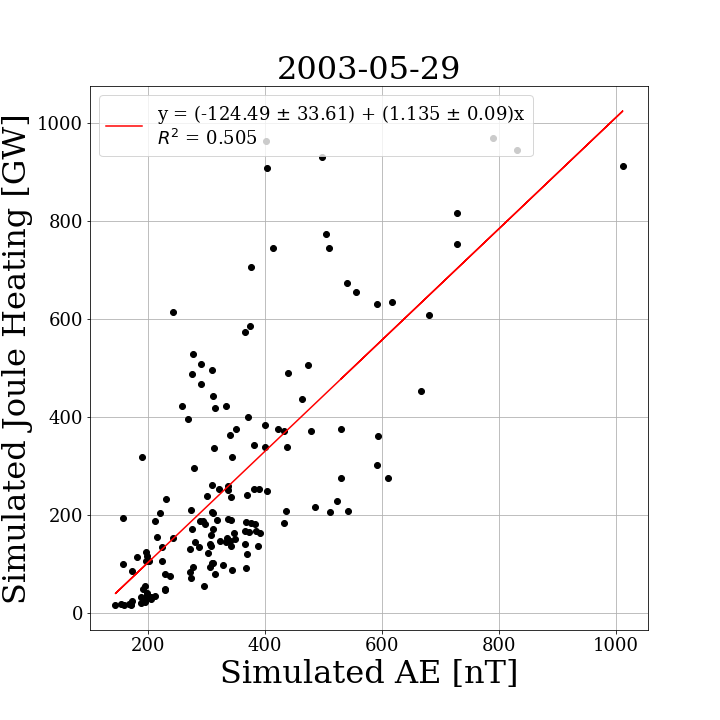}}%
{\includegraphics[clip,width=0.3\columnwidth]{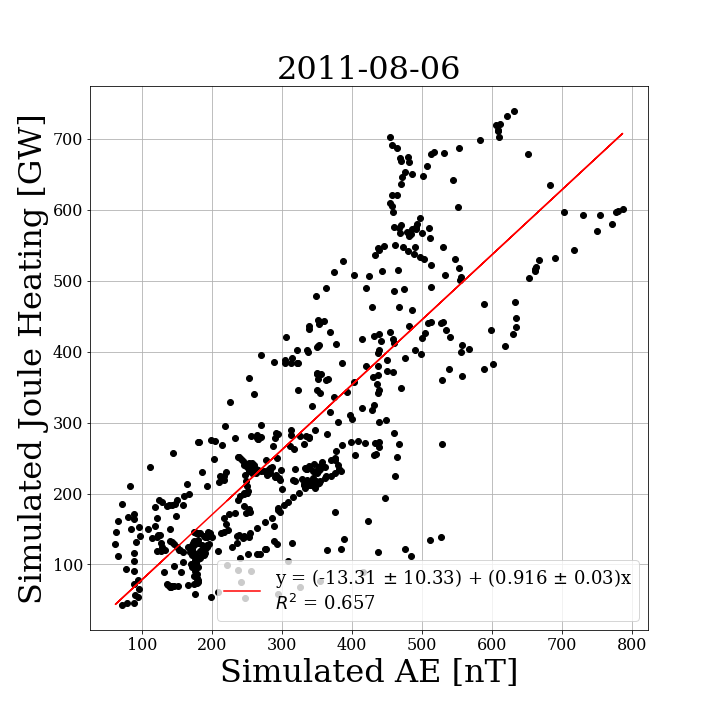}}\\%
{\includegraphics[clip,width=0.3\columnwidth]{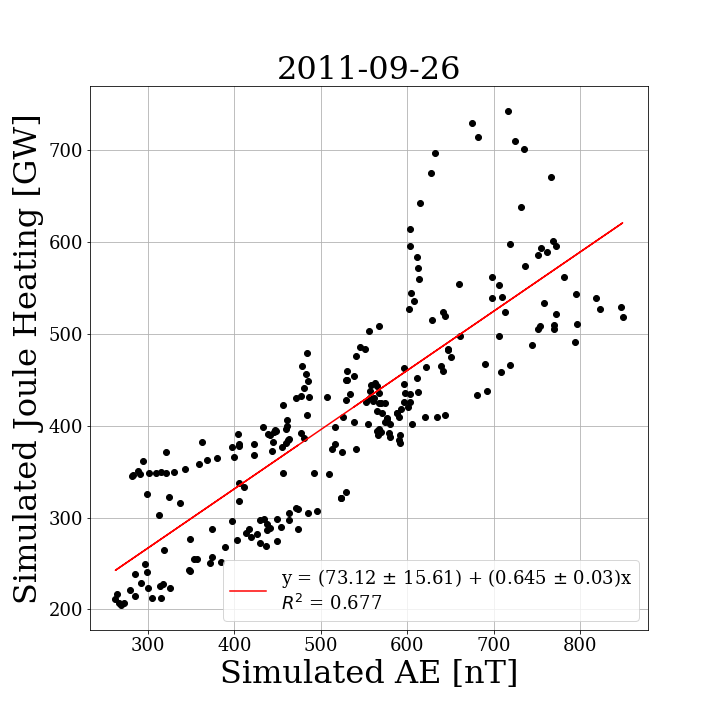} }%
{\includegraphics[clip,width=0.3\columnwidth]{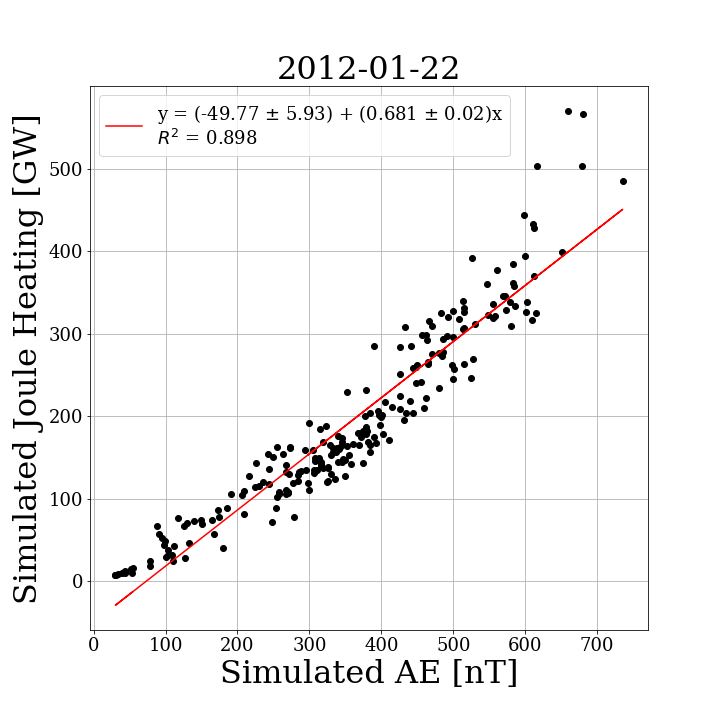}}%
{\includegraphics[clip,width=0.3\columnwidth]{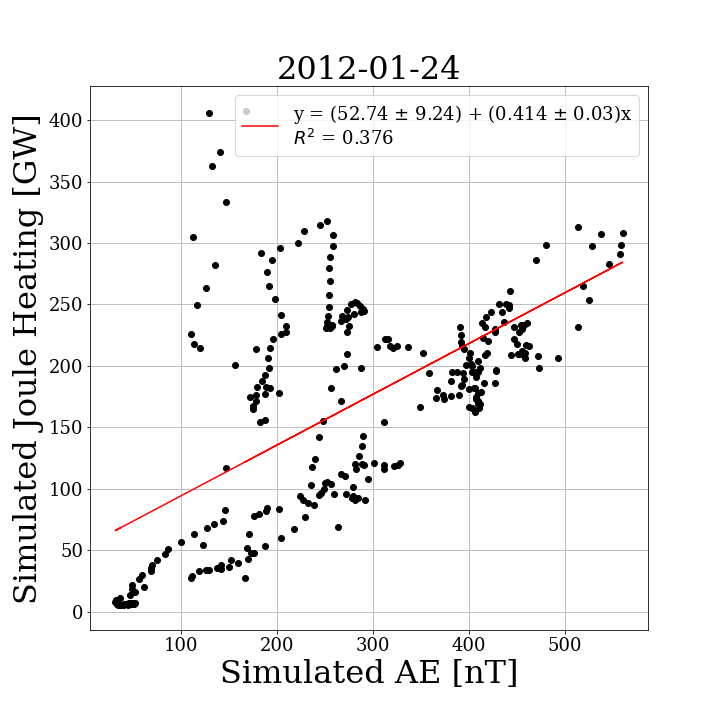} }\\%
{\includegraphics[clip,width=0.3\columnwidth]{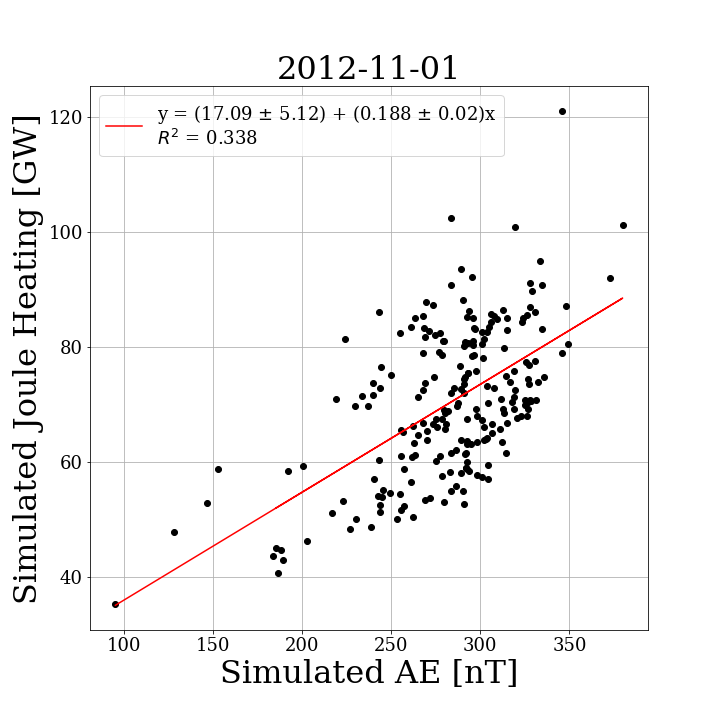}}%
{\includegraphics[clip,width=0.3\columnwidth]{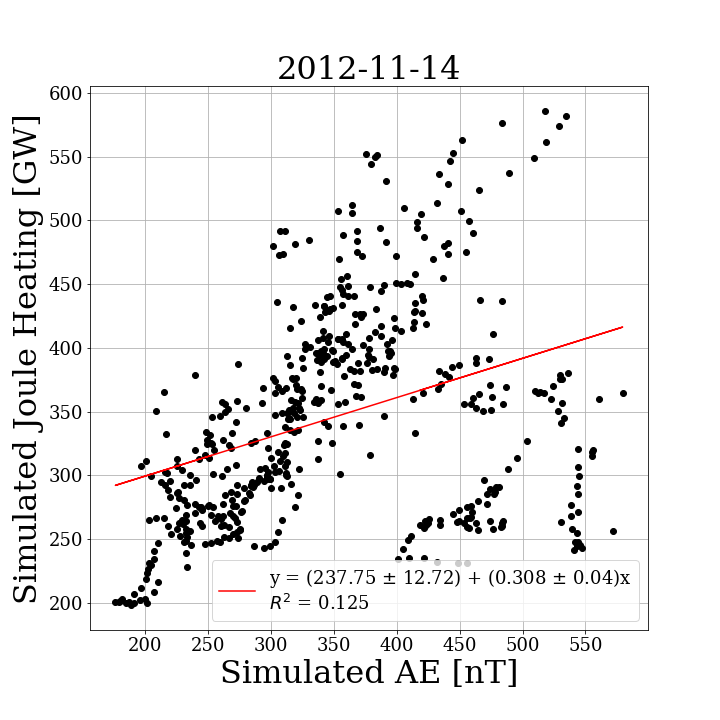} }%
{\includegraphics[clip,width=0.3\columnwidth]{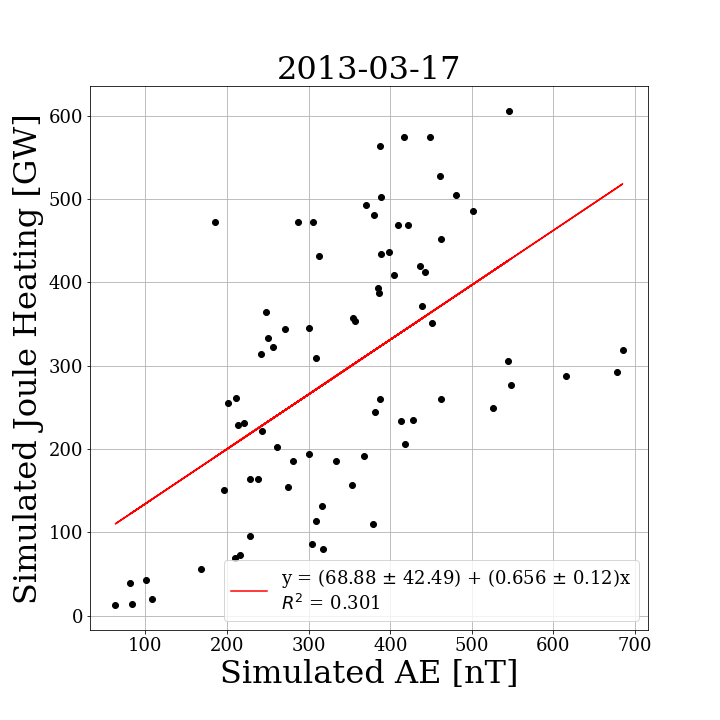}}%
\caption{Simulated Joule heating as a function of simulated AE index for the events in Table \ref{t:JHvsAE}.}
\label{fig:JHvsAE1}
\end{figure}

To investigate the relationship between Joule heating and the AE index, we use the simulations of 12 magnetic storms \citep{bagheri2022solar} with $Dst^* \leq -50$ nT. Table \ref{t:JHvsAE} summarizes the information on these events. For each storm, we find the best linear fit between simulated Joule heating and the simulated AE index. All these linear fits are shown in Figure \ref{fig:JHvsAE1}. 

\begin{table}[ht!]
\centering
\begin{tabular}{lccc}
\hline
Event Date &Time of Main Phase & Run Number&Grid\\
\hline
17 September 2000	&19-24 UT &Sean\_Blake\_042619\_1 & 1 M\\
20 April 2002	&03-07 UT &Sean\_Blake\_040519\_6& 1 M\\
23 May 2002	&11-18 UT &Luning\_Xu\_060519\_5& 1 M\\
7-8 September 2002	&16-01 UT & Sean\_Blake\_040519\_3& 1 M\\
29 May 2003	&11-24 UT&Luning\_Xu\_061419\_6& 1 M\\
5-6 August 2011	&19-04 UT&Sean\_Blake\_042619\_4& 1 M \\
26 Septemebr 2011	&15-22 UT&Pauline\_Dredger\_082321\_1& 1 M\\
22 January 2012	&07-22 UT& Diptiranjan\_Rout\_060919\_1& 1 M\\
24 January 2012	&15-20 UT&Joaquin\_Diaz\_011221\_1& 1 M\\
1 November 2012 &03-21 UT&Siyuan\_Wu\_090319\_2& 1 M\\
13-14 November 2012	&23-08 UT& Siyuan\_Wu\_120519\_2& 1 M\\
17 March 2013	&06-11 UT& Pelin\_Erdemir\_071821\_3& 1 M\\
\hline
\end{tabular}
\caption{Information on the SWMF simulations on CCMC used to study the relationship between simulated JH and the AE index. All simulations are real event simulations. For more information see Appendix A. }\label{t:JHvsAE}
\end{table}
\noindent Taking the average of all the storms' linear fits between the simulated Joule heating as a function of the simulated AE index, we can write
\begin{equation}
U_{JH} (GW) = 0.71 AE + 32~. \label{eq:NEWemipricalModel}
\end{equation}
Comparing this result to empirical models, one can conclude that the SWMF model predicts a greater dependency on the AE index for the Joule heating than the empirical models considered in the paper.

\section{Conclusion and Discussion}

In this paper, we compare the empirical models of the Joule heating with the results of SWMF simulations for a set of storm events already simulated and available at the CCMC website. We find that the SWMF simulations predict a smaller amount of Joule heating compared to empirical estimates using the SME index. Moreover, we find that events with a good correlation between the simulated ionospheric current and the AMPERE currents show a higher correlation between the simulated Joule heating and the empirical model.
\\
\\
The SWMF simulation uses the Ridley conductance model \citep{ridley2004ionospheric}. This model has two major sources of ionospheric conductance: solar EUV conductance on the dayside and auroral precipitative conductance in the polar regions. Other sources of conductance, such as seasonal dependencies, are added as functions of the dominant sources of conductance, solar zenith angle, or scalar constants \citep{mukhopadhyay2020conductance}. 
This model produces broad regions of locally elevated conductance that are discontinuous between regions of strong FACs.  On the other, the overall conductance can be lower because of unrealistically low values of particle precipitation \citep{wiltberger2012cmit}.  Therefore, during extreme events, it leads to possible unrealistic values of global quantities such as cross polar cap potential or FACs \citep{mukhopadhyay2022statistical, anderson2017comparison,liemohn2018model, welling2018recommendations}.  Consistent with these studies we find that SWMF simulations underpredict Birkeland currents for all storm events in this study.
\\
\\
To calculate Joule heating using empirical equations, we used the SME index instead of the AE index. The SME index used significantly more stations than the AE index, especially in lower latitudes. However,  the Ridley model domain in the SWMF simulation is considerably limited, from the magnetic pole to the magnetic latitude of 60° for all magnetic local times (MLTs). This could lead to inconsistency between the prediction of the SWMF simulations and empirical models for Joule heating. 
\\
\\
Two storm cases in Section 2 (13-14 November 2012 and 26 September 2011) show low correlation between the SWMF simulated Joule heating and the empirical models. To see what physical parameters are involved in the correlation coefficients, we investigated the dependency of the correlation coefficients with the solar wind parameters such as velocity, Mach number, and IMF during the main phase of the storms. Our result showed that there is not any significant evidence that these parameters can affect the correlation between SWMF and empirical models. However, increasing the magnitude of the IMF (averaged over the main phase of the storms) decreases the correlation, so there is a weak dependency on the IMF magnitude (with $R^2 = 0.26$). This result is consistent with the conclusion of \citep{welling2018recommendations}, which shows that during intense events, the ionospheric model in the SWMF simulation \citep{ridley2004ionospheric} does not predict the ionospheric conductance and indices accurately.
\\
\\
Furthermore, we investigate the relationship between the simulated Joule heating and the simulated AE index. We find that the scale factor between the AE index and the amount of Joule heating is about two times greater than in the empirical models. One possible reason for the inconsistency is that the SWMF model does not predict the AE index well \citep{haiducek2017swmf, kitamura2008properties}. The SWMF Biot-Savart calculation of the ground magnetic perturbation due to ionospheric currents does not include magnetotelluric effects. Moreover, the simulation grid latitudinal resolution is $1^{\circ}$ to $2^\circ$ (100 km to 200 km), spreading out the currents compared to reality. These effects could easily result in a factor of 2 in the calculated ground perturbation compared to observations, even if the total simulated Joule heating and current in the ionosphere is actually similar to reality.
\\
\\
The inconsistency between simulation results and empirical models is not limited to the SWMF simulation. \citet{wiltberger2012cmit} examine the Coupled Magnetosphere Ionosphere Thermosphere (CMIT) model. In CMIT, the magnetosphere model is the Lyon–Fedder–Mobarry (LFM) global magnetospheric simulation \citep{lyon2004lyon}. They found a considerable disagreement between the simulation and observational data in predicting the Cross Polar Cap Potential (CPCP) strength, hemispheric power, and SYMH index. In their study, CPCP is highly overestimated due to the weak electron precipitation power seen in the hemispheric power, leading to a low overall ionospheric conductance \citep{wiltberger2012cmit}. In addition,
\citet{pirnaris2023comparison} compared the evolution of the globally-integrated Joule heating rates between the two Global Circulation Models (GCM) of the Earth’s upper atmosphere (the Global Ionosphere/Thermosphere Model (GITM) and the Thermosphere-Ionosphere-Electrodynamics General Circulation Model (TIE-GCM)) with the several empirical models during the storm of 17 March 2015. They found that all empirical models, on average, underestimate Joule heating rates compared to both GITM and TIE-GCM, whereas TIE-GCM calculates lower heating rates compared to GITM.
\\
\\
In conclusion, there are still discrepancies between empirical models and global MHD simulations in predicting/estimating Joule heating. In this paper, we demonstrate this gap in the understanding and parametrization of Joule heating during storm times. In February 2022, 40 Space-X satellites were lost due to the enhanced Joule heating during a storm \citep{dang2022unveiling}. Therefore, a real-time prediction of Joule heating is essential for predicting the possible atmospheric drag of the satellites. Our result shows that one can use SWMF simulations for real-time prediction of Joule heating during geomagnetic storms if the SWMF result of Birkeland current is highly correlated with observations.

\section*{acknowledgements}

\noindent We gratefully acknowledge the SuperMAG website and the data provided by SuperMAG collaborators. The SME data can be found at \url{http://supermag.jhuapl.edu/indices/} for the periods described in the paper. We thank the AMPERE team and the AMPERE Science Center for providing the Iridium-derived data products. The AMPERE Birkeland current data can be found at \url{http://ampere.jhuapl.edu/products/itot/daily.html} for each day at a 2-min time resolution. We acknowledge the use of the OMNI data set, which was obtained from CDAWeb using the Space Physics Data Facility(SPDF) \url{https://cdaweb.gsfc.nasa.gov/pub/data/omni/high_res_omni/monthly_1min/}. We also gratefully acknowledge the Community Coordinated Modeling Center (CCMC) at Goddard Space Flight Center. Simulation results have been provided by the CCMC through their public Runs on Request system (\url{http://ccmc.gsfc.nasa.gov}). The CCMC is a multi-agency partnership between NASA, AFMC, AFOSR, AFRL, AFWA, NOAA, NSF, and ONR. \\
We acknowledge the support of the US National Science Foundation (NSF) under Grant No. 1916604.

\bibliographystyle{aasjournal} 
\bibliography{main}



\newpage
\appendix


\section*{Supplementary material}

# APPENDIX A: CCMC simulations used in this study

## 17 September 2000: Sean_Blake_042619_1

Run Metadata

| | |
|---|---|
| Metadata Record: | View Full Run Metadata in the CCMC Metadata Registry (CMR) |
| Model Domain: | GM |
| Model Name: | SWMF |
| Model Version: | v20180525 |
| Title/Introduction: | 2000_09_17 |
| Key Word: | 2000_09_17 |
| Run type: | Real event simulation |
| Inflow Boundary Conditions: | Time-dependent |
| Start Time: | 2000/09/17 00:00 |
| End Time: | 2000/09/19 00:00 |
| Dipole Tilt at Start in X-Z Plane: | -1.36 ° |
| Dipole Tilt in Y-Z GSE Plane: | -13.53 ° |
| Dipole Update With Time: | yes |
| Ionospheric Conductance: | auroral |
| Co-rotation: | No corotation velocity is applied at the inner boundary. |
| Grid: | 1M cells overview |
| Coordinate System for the Output: | GSM |
| Solar wind input source: | OMNI |
| Ring current model: | RCM |

Initial Solar Wind (SW) Parameters in GSM Coordinates:

| | |
|---|---|
| SW Density: | 8.60100 n/cc |
| SW Temperature: | 85661.90000 Kelvin |
| X Component of SW Velocity: | -404.30000 km/s |
| Y Component of SW Velocity: | -22.19500 km/s |
| Z Component of SW Velocity: | -36.62300 km/s |
| IMF Bx: | 0.00000 nT |
| IMF By: | 12.09100 nT |

| | |
|---|---|
| IMF Bz: | -7.35700 nT |
| IMF \|B\|: | 14.15000 nT |
| IMF Clock Angle: | 121.32000 ° |

Magnetosphere Run Parameters:

| | |
|---|---|
| GM solver: | Sokolov |
| GM limiter name: | mc3 |
| GM limiter beta: | 1.2 |
| GM implicit: | part-implicit |
| GM timestep: | 5.0 |
| GM Boris clight factor: | 0.01 |
| | |

# 20 April 2002: Sean_Blake_040519_6

Run Metadata

| | |
|---|---|
| Metadata Record: | View Full Run Metadata in the CCMC Metadata Registry (CMR) |

| | |
|---|---|
| Model Domain: | GM |
| Model Name: | SWMF |
| Model Version: | v20180525 |
| Title/Introduction: | 2002_04_20 |
| Key Word: | 2002_04_20 |

| | |
|---|---|
| Run type: | Real event simulation |
| Inflow Boundary Conditions: | Time-dependent |
| Start Time: | 2002/04/19 00:00 |
| End Time: | 2002/04/20 12:00 |
| Dipole Tilt at Start in X-Z Plane: | 8.83 ° |
| Dipole Tilt in Y-Z GSE Plane: | 30.72 ° |
| Dipole Update With Time: | yes |
| Ionospheric Conductance: | auroral |
| Co-rotation: | No corotation velocity is applied at the inner boundary. |
| Grid: | 1M cells overview |
| Coordinate System for the Output: | GSM |
| Solar wind input source: | OMNI |
| Ring current model: | RCM |

Initial Solar Wind (SW) Parameters in GSM Coordinates:

| | |
|---|---|
| SW Density: | 0.74000 n/cc |
| SW Temperature: | 15668.40000 Kelvin |
| X Component of SW Velocity: | -453.81100 km/s |
| Y Component of SW Velocity: | -2.99500 km/s |
| Z Component of SW Velocity: | -21.43800 km/s |
| IMF Bx: | 0.00000 nT |
| IMF By: | -10.58100 nT |
| IMF Bz: | -4.37800 nT |
| IMF |B|: | 11.45000 nT |
| IMF Clock Angle: | -112.48000 ° |

Magnetosphere Run Parameters:

| | |
|---|---|
| GM solver: | Sokolov |
| GM limiter name: | mc3 |
| GM limiter beta: | 1.2 |
| GM implicit: | part-implicit |
| GM timestep: | 5.0 |
| GM Boris clight factor: | 0.01 |

# 23 May 2002: Luning_Xu_060519_5

Run Metadata

| | |
|---|---|
| Metadata Record: | View Full Run Metadata in the CCMC Metadata Registry (CMR) |

| | |
|---|---|
| Model Domain: | GM |
| Model Name: | SWMF |
| Model Version: | v20180525 |
| Title/Introduction: | Event 5 |
| Key Word: | 23-May-02 |

| | |
|---|---|
| Run type: | Real event simulation |
| Inflow Boundary Conditions: | Time-dependent |
| Start Time: | 2002/05/23 04:00 |
| End Time: | 2002/05/25 21:00 |
| Dipole Tilt at Start in X-Z Plane: | 10.64 ° |
| Dipole Tilt in Y-Z GSE Plane: | 13.52 ° |
| Dipole Update With Time: | yes |
| Ionospheric Conductance: | auroral |
| Co-rotation: | No corotation velocity is applied at the inner boundary. |
| Grid: | 1M cells overview |
| Coordinate System for the Output: | GSM |
| Solar wind input source: | OMNI |
| Ring current model: | RCM |

Initial Solar Wind (SW) Parameters in GSM Coordinates:

| | |
|---|---|
| SW Density: | 17.79800 n/cc |
| SW Temperature: | 25891.30000 Kelvin |
| X Component of SW Velocity: | -402.37200 km/s |
| Y Component of SW Velocity: | 4.78500 km/s |
| Z Component of SW Velocity: | -42.79700 km/s |
| IMF Bx: | -6.21000 nT |
| IMF By: | 9.87300 nT |
| IMF Bz: | 4.14400 nT |
| IMF |B|: | 10.71000 nT |
| IMF Clock Angle: | 67.23000 ° |

Magnetosphere Run Parameters:

| | |
|---|---|
| GM solver: | Sokolov |
| GM limiter name: | mc3 |
| GM limiter beta: | 1.2 |
| GM implicit: | part-implicit |
| GM timestep: | 5.0 |
| GM Boris clight factor: | 0.01 |
| | |

# 7-8 September 2002: Sean_Blake_040519_3

Run Metadata

| | |
|---|---|
| Metadata Record: | View Full Run Metadata in the CCMC Metadata Registry (CMR) |

| | |
|---|---|
| Model Domain: | GM |
| Model Name: | SWMF |
| Model Version: | v20180525 |
| Title/Introduction: | 2002_09_07 |
| Key Word: | 2002_09_07 |

| | |
|---|---|
| Run type: | Real event simulation |
| Inflow Boundary Conditions: | Time-dependent |
| Start Time: | 2002/09/07 12:00 |
| End Time: | 2002/09/08 12:00 |
| Dipole Tilt at Start in X-Z Plane: | 10.95 ° |
| Dipole Tilt in Y-Z GSE Plane: | -32.61 ° |
| Dipole Update With Time: | yes |
| Ionospheric Conductance: | auroral |
| Co-rotation: | No corotation velocity is applied at the inner boundary. |
| Grid: | 1M cells overview |
| Coordinate System for the Output: | GSM |
| Solar wind input source: | OMNI |
| Ring current model: | RCM |

Initial Solar Wind (SW) Parameters in GSM Coordinates:

| | |
|---|---|
| SW Density: | 5.14500 n/cc |
| SW Temperature: | 12737.60000 Kelvin |
| X Component of SW Velocity: | -355.54000 km/s |
| Y Component of SW Velocity: | -25.42300 km/s |
| Z Component of SW Velocity: | -3.82700 km/s |
| IMF Bx: | 0.00000 nT |
| IMF By: | 3.93100 nT |
| IMF Bz: | -3.63600 nT |
| IMF |B|: | 5.35000 nT |
| IMF Clock Angle: | 132.77000 ° |

Magnetosphere Run Parameters:

| | |
|---|---|
| GM solver: | Sokolov |
| GM limiter name: | mc3 |
| GM limiter beta: | 1.2 |
| GM implicit: | part-implicit |
| GM timestep: | 5.0 |
| GM Boris clight factor: | 0.01 |
| | |

# 29 May 2003: Luning_Xu_061419_6

Run Metadata

| | |
|---|---|
| Metadata Record: | View Full Run Metadata in the CCMC Metadata Registry (CMR) |
| Model Domain: | GM |
| Model Name: | SWMF |
| Model Version: | v20180525 |
| Title/Introduction: | Event 6 |
| Key Word: | 29-May-03 |
| Run type: | Real event simulation |
| Inflow Boundary Conditions: | Time-dependent |
| Start Time: | 2003/05/29 05:00 |
| End Time: | 2003/05/31 08:00 |
| Dipole Tilt at Start in X-Z Plane: | 11.39 ° |
| Dipole Tilt in Y-Z GSE Plane: | 8.72 ° |
| Dipole Update With Time: | yes |
| Ionospheric Conductance: | auroral |
| Co-rotation: | No corotation velocity is applied at the inner boundary. |
| Grid: | 1M cells overview |
| Coordinate System for the Output: | GSM |
| Solar wind input source: | OMNI |
| Ring current model: | RCM |

Initial Solar Wind (SW) Parameters in GSM Coordinates:

| | |
|---|---|
| SW Density: | 1.85000 n/cc |
| SW Temperature: | 93325.70000 Kelvin |
| X Component of SW Velocity: | -622.92200 km/s |
| Y Component of SW Velocity: | -1.93800 km/s |
| Z Component of SW Velocity: | 34.14000 km/s |
| IMF Bx: | 2.18000 nT |
| IMF By: | -2.86200 nT |
| IMF Bz: | -0.41700 nT |
| IMF |B|: | 2.89000 nT |
| IMF Clock Angle: | -98.29000 ° |

Magnetosphere Run Parameters:

| | |
|---|---|
| GM solver: | Sokolov |
| GM limiter name: | mc3 |
| GM limiter beta: | 1.2 |
| GM implicit: | part-implicit |
| GM timestep: | 5.0 |
| GM Boris clight factor: | 0.01 |
| | |

# 2 May 2010: Pelin_Erdemir_021419_1

Run Metadata

| | |
|---|---|
| Metadata Record: | View Full Run Metadata in the CCMC Metadata Registry (CMR) |

| | |
|---|---|
| Model Domain: | GM |
| Model Name: | SWMF |
| Model Version: | v20180525 |
| Title/Introduction: | Joule Heating |
| Key Word: | Joule Heating |

| | |
|---|---|
| Run type: | Real event simulation |
| Inflow Boundary Conditions: | Time-dependent |
| Start Time: | 2010/05/02 00:00 |
| End Time: | 2010/05/04 00:00 |
| Dipole Tilt at Start in X-Z Plane: | 13.41 ° |
| Dipole Tilt in Y-Z GSE Plane: | 27.64 ° |
| Dipole Update With Time: | yes |
| Ionospheric Conductance: | auroral |
| Co-rotation: | No corotation velocity is applied at the inner boundary. |
| Grid: | 1M cells overview |
| Coordinate System for the Output: | GSM |
| Solar wind input source: | WIND |
| Ring current model: | CRCM |

Initial Solar Wind (SW) Parameters in GSM Coordinates:

| | |
|---|---|
| SW Density: | 7.59100 n/cc |
| SW Temperature: | 25170.30000 Kelvin |
| X Component of SW Velocity: | -317.23300 km/s |
| Y Component of SW Velocity: | -30.42000 km/s |
| Z Component of SW Velocity: | -13.65200 km/s |
| IMF Bx: | 0.00000 nT |
| IMF By: | 2.10800 nT |
| IMF Bz: | 1.25600 nT |
| IMF |B|: | 2.45000 nT |
| IMF Clock Angle: | 59.21000 ° |

Magnetosphere Run Parameters:

| | |
|---|---|
| GM solver: | Sokolov |
| GM limiter name: | mc3 |
| GM limiter beta: | 1.2 |
| GM implicit: | part-implicit |
| GM timestep: | 5.0 |
| GM Boris clight factor: | 0.01 |
| | |

# 5-6 August 2011: Sean_Blake_042619_4

Run Metadata

| | |
|---|---|
| Metadata Record: | View Full Run Metadata in the CCMC Metadata Registry (CMR) |

| | |
|---|---|
| Model Domain: | GM |
| Model Name: | SWMF |
| Model Version: | v20180525 |
| Title/Introduction: | 2011_08_05 |
| Key Word: | 2011_08_05 |

| | |
|---|---|
| Run type: | Real event simulation |
| Inflow Boundary Conditions: | Time-dependent |
| Start Time: | 2011/08/05 00:00 |
| End Time: | 2011/08/07 00:00 |
| Dipole Tilt at Start in X-Z Plane: | 14.23 ° |
| Dipole Tilt in Y-Z GSE Plane: | -6.46 ° |
| Dipole Update With Time: | yes |
| Ionospheric Conductance: | auroral |
| Co-rotation: | No corotation velocity is applied at the inner boundary. |
| Grid: | 1M cells overview |
| Coordinate System for the Output: | GSM |
| Solar wind input source: | OMNI |
| Ring current model: | RCM |

Initial Solar Wind (SW) Parameters in GSM Coordinates:

| | |
|---|---|
| SW Density: | 4.75000 n/cc |
| SW Temperature: | 170137.00000 Kelvin |
| X Component of SW Velocity: | -410.00000 km/s |
| Y Component of SW Velocity: | -34.47500 km/s |
| Z Component of SW Velocity: | -91.87500 km/s |
| IMF Bx: | 0.00000 nT |
| IMF By: | -0.37600 nT |
| IMF Bz: | -7.43300 nT |
| IMF |B|: | 7.44000 nT |
| IMF Clock Angle: | -177.10000 ° |

Magnetosphere Run Parameters:

| | |
|---|---|
| GM solver: | Sokolov |
| GM limiter name: | mc3 |
| GM limiter beta: | 1.2 |
| GM implicit: | part-implicit |
| GM timestep: | 5.0 |
| GM Boris clight factor: | 0.01 |
| | |

# 26 September 2011: Pauline_Dredger_082321_1

Run Metadata

| | |
|---|---|
| Metadata Record: | View Full Run Metadata in the CCMC Metadata Registry (CMR) |
| Model Domain: | GM |
| Model Name: | SWMF |
| Model Version: | v20180525 |
| Title/Introduction: | Magnetopause Motion |
| Key Word: | magnetopause ring current |
| Run type: | Real event simulation |
| Inflow Boundary Conditions: | Time-dependent |
| Start Time: | 2011/09/26 07:00 |
| End Time: | 2011/09/26 19:00 |
| Dipole Tilt at Start in X-Z Plane: | -10.54 ° |
| Dipole Tilt in Y-Z GSE Plane: | -29.10 ° |
| Dipole Update With Time: | yes |
| Ionospheric Conductance: | auroral |
| Co-rotation: | No corotation velocity is applied at the inner boundary. |
| Grid: | 1M cells overview |
| Coordinate System for the Output: | GSM |
| Solar wind input source: | OMNI |
| Ring current model: | RCM |

Initial Solar Wind (SW) Parameters in GSM Coordinates:

| | |
|---|---|
| SW Density: | 13.38500 n/cc |
| SW Temperature: | 74214.70000 Kelvin |
| X Component of SW Velocity: | -340.43200 km/s |
| Y Component of SW Velocity: | -4.08100 km/s |
| Z Component of SW Velocity: | 13.66100 km/s |
| IMF Bx: | 0.00000 nT |
| IMF By: | 1.40200 nT |
| IMF Bz: | 6.42900 nT |
| IMF |B|: | 6.58000 nT |
| IMF Clock Angle: | 12.30000 ° |

Magnetosphere Run Parameters:

| | |
|---|---|
| GM solver: | Sokolov |
| GM limiter name: | mc3 |
| GM limiter beta: | 1.2 |
| GM implicit: | part-implicit |
| GM timestep: | 2.0 |
| GM Boris clight factor: | 0.01 |
| | |

# 22 January 2012: Diptiranjan_Rout_060919_1

Run Metadata

| | |
|---|---|
| Metadata Record: | View Full Run Metadata in the CCMC Metadata Registry (CMR) |
| Model Domain: | GM |
| Model Name: | SWMF |
| Model Version: | v20180525 |
| Title/Introduction: | Solar wind pressure |
| Key Word: | FAC and Electric field |
| Run type: | Real event simulation |
| Inflow Boundary Conditions: | Time-dependent |
| Start Time: | 2012/01/22 00:00 |
| End Time: | 2012/01/22 23:44 |
| Dipole Tilt at Start in X-Z Plane: | -23.78 ° |
| Dipole Tilt in Y-Z GSE Plane: | 23.00 ° |
| Dipole Update With Time: | yes |
| Ionospheric Conductance: | auroral |
| Co-rotation: | No corotation velocity is applied at the inner boundary. |
| Grid: | 1M cells overview |
| Coordinate System for the Output: | GSM |
| Solar wind input source: | OMNI |

Initial Solar Wind (SW) Parameters in GSM Coordinates:

| | |
|---|---|
| SW Density: | 12.17700 n/cc |
| SW Temperature: | 14859.60000 Kelvin |
| X Component of SW Velocity: | -331.78300 km/s |
| Y Component of SW Velocity: | -6.09500 km/s |
| Z Component of SW Velocity: | -17.36900 km/s |
| IMF Bx: | 0.00000 nT |
| IMF By: | -8.38000 nT |
| IMF Bz: | 8.72000 nT |
| IMF |B|: | 12.09000 nT |
| IMF Clock Angle: | -43.86000 ° |

Magnetosphere Run Parameters:

| | |
|---|---|
| GM solver: | Sokolov |
| GM limiter name: | mc3 |
| GM limiter beta: | 1.2 |
| GM implicit: | part-implicit |
| GM timestep: | 5.0 |
| GM Boris clight factor: | 0.01 |
| | |

# 24 January 2012: Joaquin_Diaz_011221_1

Run Metadata

| | |
|---|---|
| Metadata Record: | View Full Run Metadata in the CCMC Metadata Registry (CMR) |
| Model Domain: | GM |
| Model Name: | SWMF |
| Model Version: | v20180525 |
| Title/Introduction: | Batsrus run for Jan2012 |
| Key Word: | Batsrus with RCM |
| Run type: | Real event simulation |
| Inflow Boundary Conditions: | Time-dependent |
| Start Time: | 2012/01/24 00:00 |
| End Time: | 2012/01/24 23:59 |
| Dipole Tilt at Start in X-Z Plane: | -23.38 ° |
| Dipole Tilt in Y-Z GSE Plane: | 23.68 ° |
| Dipole Update With Time: | yes |
| Ionospheric Conductance: | auroral |
| Co-rotation: | No corotation velocity is applied at the inner boundary. |
| Grid: | high-resolution grid with 9,623,552 cells |
| Coordinate System for the Output: | GSM |
| Solar wind input source: | OMNI |
| Ring current model: | RCM |

Initial Solar Wind (SW) Parameters in GSM Coordinates:

| | |
|---|---|
| SW Density: | 3.29000 n/cc |
| SW Temperature: | 16312.00000 Kelvin |
| X Component of SW Velocity: | -403.40000 km/s |
| Y Component of SW Velocity: | -12.66200 km/s |
| Z Component of SW Velocity: | 8.97600 km/s |
| IMF Bx: | 0.00000 nT |
| IMF By: | -3.27000 nT |
| IMF Bz: | -5.74000 nT |
| IMF |B|: | 6.61000 nT |
| IMF Clock Angle: | -150.33000 ° |

Magnetosphere Run Parameters:

| | |
|---|---|
| GM solver: | Rusanov |
| GM limiter name: | mc3 |
| GM limiter beta: | 1 |
| GM implicit: | explicit |
| GM Boris clight factor: | 0.01 |

# 15 July 2012: Antti_Lakka_070918_1

Run Metadata

| | |
|---|---|
| Metadata Record: | View Full Run Metadata in the CCMC Metadata Registry (CMR) |
| Model Domain: | GM |
| Model Name: | SWMF |
| Model Version: | v20140611 |
| Title/Introduction: | July 2012 ICME |
| Key Word: | ICME |
| Run type: | Real event simulation |
| Inflow Boundary Conditions: | Time-dependent |
| Start Time: | 2012/07/14 07:00 |
| End Time: | 2012/07/17 15:00 |
| Dipole Tilt at Start in X-Z Plane: | 13.52 ° |
| Dipole Tilt in Y-Z GSE Plane: | -14.52 ° |
| Dipole Update With Time: | yes |
| Ionospheric Conductance: | auroral |
| Co-rotation: | No corotation velocity is applied at the inner boundary. |
| Grid: | 1M cells overview |
| Coordinate System for the Output: | GSM |
| Solar wind input source: | OMNI |
| Ring current model: | RCM |

Initial Solar Wind (SW) Parameters in GSM Coordinates:

| | |
|---|---|
| SW Density: | 4.62400 n/cc |
| SW Temperature: | 11841.10000 Kelvin |
| X Component of SW Velocity: | -316.57300 km/s |
| Y Component of SW Velocity: | 8.71300 km/s |
| Z Component of SW Velocity: | 3.40200 km/s |
| IMF Bx: | 0.00000 nT |
| IMF By: | -1.97300 nT |
| IMF Bz: | -0.01100 nT |
| IMF |B|: | 1.97000 nT |
| IMF Clock Angle: | -90.32000 ° |

Magnetosphere Run Parameters:

| | |
|---|---|
| GM solver: | Sokolov |
| GM limiter name: | mc3 |
| GM limiter beta: | 1.2 |
| GM implicit: | part-implicit |
| GM timestep: | 5.0 |
| GM Boris clight factor: | 0.01 |
| | |

# 8-9 October 2012: Sean_Blake_042619_7

Run Metadata

| | |
|---|---|
| Metadata Record: | View Full Run Metadata in the CCMC Metadata Registry (CMR) |

| | |
|---|---|
| Model Domain: | GM |
| Model Name: | SWMF |
| Model Version: | v20180525 |
| Title/Introduction: | 2012_10_08 |
| Key Word: | 2012_10_08 |

| | |
|---|---|
| Run type: | Real event simulation |
| Inflow Boundary Conditions: | Time-dependent |
| Start Time: | 2012/10/07 20:00 |
| End Time: | 2012/10/09 06:00 |
| Dipole Tilt at Start in X-Z Plane: | 0.38 ° |
| Dipole Tilt in Y-Z GSE Plane: | -15.15 ° |
| Dipole Update With Time: | yes |
| Ionospheric Conductance: | auroral |
| Co-rotation: | No corotation velocity is applied at the inner boundary. |
| Grid: | 1M cells overview |
| Coordinate System for the Output: | GSM |
| Solar wind input source: | OMNI |
| Ring current model: | RCM |

Initial Solar Wind (SW) Parameters in GSM Coordinates:

| | |
|---|---|
| SW Density: | 4.32900 n/cc |
| SW Temperature: | 17689.10000 Kelvin |
| X Component of SW Velocity: | -329.49800 km/s |
| Y Component of SW Velocity: | -12.20300 km/s |
| Z Component of SW Velocity: | 2.77000 km/s |
| IMF Bx: | 0.00000 nT |
| IMF By: | 4.87100 nT |
| IMF Bz: | -5.96700 nT |
| IMF |B|: | 7.70000 nT |
| IMF Clock Angle: | 140.77000 ° |

Magnetosphere Run Parameters:

| | |
|---|---|
| GM solver: | Sokolov |
| GM limiter name: | mc3 |
| GM limiter beta: | 1.2 |
| GM implicit: | part-implicit |
| GM timestep: | 5.0 |
| GM Boris clight factor: | 0.01 |
| | |

# 1 November 2012: Siyuan_Wu_090319_2

Run Metadata

| | |
|---|---|
| Metadata Record: | View Full Run Metadata in the CCMC Metadata Registry (CMR) |

| | |
|---|---|
| Model Domain: | GM |
| Model Name: | SWMF |
| Model Version: | v20180525 |
| Title/Introduction: | Four day duration |
| Key Word: | Storm20121101 |

| | |
|---|---|
| Run type: | Real event simulation |
| Inflow Boundary Conditions: | Time-dependent |
| Start Time: | 2012/10/31 00:00 |
| End Time: | 2012/11/03 00:00 |
| Dipole Tilt at Start in X-Z Plane: | -17.80 ° |
| Dipole Tilt in Y-Z GSE Plane: | -9.30 ° |
| Dipole Update With Time: | yes |
| Ionospheric Conductance: | auroral |
| Co-rotation: | No corotation velocity is applied at the inner boundary. |
| Grid: | 2M cells |
| Coordinate System for the Output: | GSM |
| Solar wind input source: | OMNI |
| Ring current model: | RCM |

Initial Solar Wind (SW) Parameters in GSM Coordinates:

| | |
|---|---|
| SW Density: | 8.05200 n/cc |
| SW Temperature: | 40781.60000 Kelvin |
| X Component of SW Velocity: | -280.62500 km/s |
| Y Component of SW Velocity: | 16.22800 km/s |
| Z Component of SW Velocity: | 12.14900 km/s |
| IMF Bx: | 0.79000 nT |
| IMF By: | 3.38700 nT |
| IMF Bz: | 1.98900 nT |
| IMF |B|: | 3.93000 nT |
| IMF Clock Angle: | 59.58000 ° |

Magnetosphere Run Parameters:

| | |
|---|---|
| GM solver: | Rusanov |
| GM limiter name: | mc3 |
| GM limiter beta: | 1 |
| GM implicit: | explicit |
| GM Boris clight factor: | 0.01 |

# 13-14 November 2012: Siyuan_Wu_120519_2

Run Metadata

| | |
|---|---|
| Metadata Record: | View Full Run Metadata in the CCMC Metadata Registry (CMR) |

| | |
|---|---|
| Model Domain: | GM |
| Model Name: | SWMF |
| Model Version: | v20180525 |
| Title/Introduction: | Storm_20121112 |
| Key Word: | Storm_at_CNH and same long and lat |

| | |
|---|---|
| Run type: | Real event simulation |
| Inflow Boundary Conditions: | Time-dependent |
| Start Time: | 2012/11/12 00:00 |
| End Time: | 2012/11/15 00:00 |
| Dipole Tilt at Start in X-Z Plane: | -21.19 ° |
| Dipole Tilt in Y-Z GSE Plane: | -5.78 ° |
| Dipole Update With Time: | yes |
| Ionospheric Conductance: | auroral |
| Co-rotation: | No corotation velocity is applied at the inner boundary. |
| Grid: | 1M cells overview |
| Coordinate System for the Output: | GSM |
| Solar wind input source: | OMNI |
| Ring current model: | RCM |

Initial Solar Wind (SW) Parameters in GSM Coordinates:

| | |
|---|---|
| SW Density: | 6.21600 n/cc |
| SW Temperature: | 84174.10000 Kelvin |
| X Component of SW Velocity: | -308.00000 km/s |
| Y Component of SW Velocity: | 21.18600 km/s |
| Z Component of SW Velocity: | -10.66000 km/s |
| IMF Bx: | -0.17000 nT |
| IMF By: | -1.53200 nT |
| IMF Bz: | -1.61400 nT |
| IMF |B|: | 2.23000 nT |

| | |
|---|---|
| IMF Clock Angle: | -136.49000 ° |

Magnetosphere Run Parameters:

| | |
|---|---|
| GM solver: | Sokolov |
| GM limiter name: | mc3 |
| GM limiter beta: | 1.2 |
| GM implicit: | part-implicit |
| GM timestep: | 5.0 |
| GM Boris clight factor: | 0.01 |
| | |

# 17 March 2013: Pelin_Erdemir_071821_3

Run Status: Run Complete

Status updated: 2021-07-26T23:32:03+0000

## Run Metadata

| | |
|---|---|
| Metadata Record: | View Full Run Metadata in the CCMC Metadata Registry (CMR) |
| Model Domain: | GM |
| Model Name: | SWMF |
| Model Version: | v20180525 |
| Title/Introduction: | Joule Heating |
| Key Word: | Joule Heating |
| Run type: | Real event simulation |
| Inflow Boundary Conditions: | Time-dependent |
| Start Time: | 2013/03/16 00:00 |
| End Time: | 2013/03/22 00:00 |
| Dipole Tilt at Start in X-Z Plane: | -5.14 ° |
| Dipole Tilt in Y-Z GSE Plane: | 32.84 ° |
| Dipole Update With Time: | yes |
| Ionospheric Conductance: | auroral |
| Co-rotation: | No corotation velocity is applied at the inner boundary. |
| Grid: | 1M cells overview |
| Coordinate System for the Output: | GSM |
| Solar wind input source: | WIND |
| Ring current model: | RCM |

Initial Solar Wind (SW) Parameters in GSM Coordinates:

| | |
|---|---|
| SW Density: | 3.42500 n/cc |
| SW Temperature: | 67113.60000 Kelvin |
| X Component of SW Velocity: | -441.77300 km/s |
| Y Component of SW Velocity: | -5.03600 km/s |
| Z Component of SW Velocity: | 21.89300 km/s |
| IMF Bx: | 0.00000 nT |
| IMF By: | 3.69200 nT |
| IMF Bz: | -2.17100 nT |

| | |
|---|---|
| IMF \|B\|: | 4.28000 nT |
| IMF Clock Angle: | 120.46000 ° |

Magnetosphere Run Parameters:

| | |
|---|---|
| GM solver: | Sokolov |
| GM limiter name: | mc3 |
| GM limiter beta: | 1.2 |
| GM implicit: | part-implicit |
| GM timestep: | 2.0 |
| GM Boris clight factor: | 0.01 |

# 9 September 2015: Shannon_Hill_061518_1

Run Metadata

| | |
|---|---|
| Metadata Record: | View Full Run Metadata in the CCMC Metadata Registry (CMR) |

| | |
|---|---|
| Model Domain: | GM |
| Model Name: | SWMF |
| Model Version: | v20130129 |
| Title/Introduction: | Sept 2015 Storm |
| Key Word: | storm |

| | |
|---|---|
| Run type: | Real event simulation |
| Inflow Boundary Conditions: | Time-dependent |
| Start Time: | 2015/09/07 00:00 |
| End Time: | 2015/09/11 00:00 |
| Dipole Tilt at Start in X-Z Plane: | 3.35 ° |
| Dipole Tilt in Y-Z GSE Plane: | -13.43 ° |
| Dipole Update With Time: | yes |
| Ionospheric Conductance: | auroral |
| Co-rotation: | No corotation velocity is applied at the inner boundary. |
| Grid: | 2M cells |
| Coordinate System for the Output: | GSM |
| Solar wind input source: | OMNI |
| Ring current model: | CRCM |

Initial Solar Wind (SW) Parameters in GSM Coordinates:

| | |
|---|---|
| SW Density: | 4.90600 n/cc |
| SW Temperature: | 124576.00000 Kelvin |
| X Component of SW Velocity: | -477.71700 km/s |
| Y Component of SW Velocity: | -43.23000 km/s |
| Z Component of SW Velocity: | -6.58600 km/s |
| IMF Bx: | 1.79000 nT |
| IMF By: | 6.04000 nT |
| IMF Bz: | 3.16500 nT |
| IMF |B|: | 6.82000 nT |
| IMF Clock Angle: | 62.35000 ° |

Magnetosphere Run Parameters:

| | |
|---|---|
| GM solver: | Rusanov |
| GM limiter name: | mc3 |
| GM limiter beta: | 1 |
| GM implicit: | explicit |
| GM Boris clight factor: | 0.01 |



\end{document}